\documentstyle[preprint,eqsecnum,aps]{revtex}
\tightenlines
\begin{document}
\draft
\title{Nonrelativistic Limit of the Scalar Chern-Simons Theory and the
Aharonov-Bohm Scattering}
\author{M. Gomes$^{\,a}$, J. M. C. Malbouisson$^{\,b}$ and A. J. da 
Silva$^{\,a}$}
\address{$^{a\,}$Instituto de F\'\i sica, Universidade de S\~ao Paulo, 
Caixa Postal 66318,\\
05315--970, S\~ao Paulo, SP, Brazil.}
\address{$^{b\,}$Instituto de F\'\i sica, Universidade Federal da Bahia,\\
40210--340, Salvador, BA, Brazil.}

\maketitle

\begin{abstract}
We study the nonrelativistic limit of the quantum theory of a Chern-Simons
field minimally coupled to a scalar field with quartic self-interaction. The
renormalization of the relativistic model, in the Coulomb gauge, is
discussed. We employ a procedure to calculate scattering amplitudes for
low momenta that generates their $|{\bf p}|/m$ expansion and separates the
contributions coming from high and low energy intermediary states. The two
body scattering amplitude is calculated up to order ${\bf p}^2/m^2$. It is
shown that the existence of a critical value of the self-interaction
parameter for which the 2-particle scattering amplitude reduces to the
Aharonov-Bohm one is a strictly nonrelativistic feature. The subdominant terms 
correspond to relativistic corrections to the Aharonov-Bohm scattering. A
nonrelativistic reduction scheme and an effective nonrelativistic
Lagrangian to account for the relativistic corrections are proposed.
\end{abstract}

\narrowtext

\section{INTRODUCTION}

It is generally expected that nonrelativistic quantum theories, which
provide very good descriptions of many physical phenomena, might be obtained
from corresponding relativistic ones as appropriate limits for low momenta.
In relativistic quantum mechanics, the reduction to nonrelativistic wave
equations and Hamiltonians is based on the use of canonical transformations
of Foldy-Wouthuysen type \cite{fol}. For a relativistic field theory,
at classical level, such a procedure can be applied to the bilinear part of
the Lagrangian but the treatment of interactions is not straightforward.
Another way to arrive at Schr\"odinger type of Lagrangian and equations of
motion is to redefine the fields, by separating the factor $\exp
(-imc^2t/\hbar )$, and then take the large $c$ (or large-mass) limit which
eliminates rapidly oscillating terms \cite{beg}. The nonrelativistic
Chern-Simons (CS) Lagrangian, which is relevant for the physics of anyons 
\cite{lm}, can be obtained from the relativistic one by performing such a
limit \cite{jac}. Due to quantum oscillations and renormalization
problems, however, the nonrelativistic limit of a quantum field theory is
much more subtle. In this respect, inspired in renormalization group
arguments, there has been proposals of construction of effective Lagrangians
by amending the nonrelativistic theories with other interaction terms,
representing the effect of the integration of the relativistic degrees of
freedom \cite{lep}.

Recently \cite{go}, we have introduced a scheme of nonrelativistic
approximation of a quantum field theory which uses an intermediate
cutoff device and allows a systematic expansion
of the scattering amplitudes in powers of $|{\bf p}|/m$, permitting the
identification, in the Hilbert space of intermediate states, of the origin
of each contribution to the S-matrix perturbative series. The procedure was
applied to $\lambda \phi ^4$ theory in 2+1 dimensions, where there exist
exact results in one loop order, so that it could be explicitly verified.
This model has a well defined nonrelativistic counterpart which presents an
interesting scale anomaly \cite{ber}.

Here, we extend the discussion of nonrelativistic limit to a more
involving and physically appealing gauge theory by considering the
Chern-Simons field minimally coupled to a scalar field with quartic
self-interaction \cite{Na}. The coupling of a matter field with a
gauge field governed by a Chern-Simons action \cite{hag} has also
been studied as a limiting case of the
topologically massive gauge theory \cite{djt} in a covariant gauge 
\cite{sem}, and more recently, the quantization of the fermionic model,
in the Coulomb gauge, has been constructed \cite{fle}. The
corresponding nonrelativistic model \cite{hag2}, including the quartic
self-interaction \cite{jac,bl}, has been suggested as a
field-theoretical formulation of the Aharonov-Bohm (AB) scattering 
\cite{ab} and as an effective theory for the fractional quantum Hall
effect \cite{fra}.

In the nonrelativistic theory, there exist critical values of the
renormalized self-coupling parameter for which the one loop scattering
vanishes and scale invariance is maintained. By choosing the positive
coupling, corresponding to a repulsive contact interaction, the tree level
reduces to the AB scattering \cite{bl}. The existence of this scale
invariance at the critical self-interaction has been explicitly verified up
to three loops using differential regularization \cite{flr} and was
recently proven to hold in all orders of perturbation theory \cite{kim}. 
This has also been obtained for the nonrelativistic limit (in leading
order) of the 1-loop particle-particle scattering of scalar self-interacting
particles minimally coupled to a CS gauge field \cite{boz}. This
criticality, however, ceases to exist in the relativistic case \cite{go2}.

In this article we discuss the nonrelativistic approximation of the
scattering amplitudes in the scalar Chern-Simons theory by using the
above mentioned cutoff procedure.
After computing next to leading nonrelativistic contributions to the two
body scattering, we are able to construct a Galilean effective Lagrangian
which extends the Aharonov-Bohm theory.
We organize the article as follows. In Sec. II, we present the relativistic
model and discuss its renormalization at one loop level. We then, in Sec.
III, introduce the intermediate cutoff procedure we use to obtain the $|{\bf %
p}|/m$ expansion of the quantum amplitudes. The calculation of the 1-loop
particle-particle scattering amplitude up to order ${\bf p}^2/m^2$ is
presented in Sec. IV and in the Appendix. It is seen that the leading term
of the $|{\bf p}|/m$ expansion possesses the same critical self-coupling as
in the nonrelativistic case. However, the subdominant parts do not vanish at
the critical self-interaction values and so the aforesaid criticality is
strictly nonrelativistic. The implications of this fact for the AB
scattering are discussed in Sec. V where we also introduce a nonrelativistic
reduction scheme for the scattering amplitudes. We consider the possibility
of generating the relativistic corrections by adding nonrenormalizable
interactions to the nonrelativistic Lagrangian and suggest an effective
Lagrangian that accounts for the results up to order ${\bf p}^2/m^2$ .

\section{The Relativistic Model}

We consider a charged self-interacting scalar field in 2 + 1 dimensions
minimally coupled to a Chern-Simons gauge field described by the Lagrangian
density

\begin{equation}
{\cal L}=(D_\mu \phi )^{*}(D^\mu \phi )-m^2\phi ^{*}\phi -\frac \lambda 4%
(\phi ^{*}\phi )^2+\frac \Theta 2\epsilon _{\sigma \mu \nu }A^\sigma
\partial ^\mu A^\nu ,  \label{lagran}
\end{equation}
\noindent where $D_\mu =\partial _\mu +ieA_\mu $ is the covariant
derivative, $\epsilon _{\sigma \mu \nu }$ is the fully antisymmetric tensor
normalized to $\epsilon _{012}=+1$, the Minkowski metric signature is (1,
-1, -1) and the units are such that $\hbar =c=1$.

As it stands, the $A_\mu $ is not a dynamical variable, its equation of
motion represents a constraint and so there does not exist real gauge
particles \cite{hag}. However, one has to specify a gauge to work with
and, by adding a gauge fixing term to the Lagrangian (\ref{lagran}), one can
treat the gauge field dynamically in the internal lines. For convenience in
discussing the nonrelativistic limit, we choose the Coulomb gauge where

\begin{equation}
{\cal L}_{GF}=-\frac \xi 2(\partial _iA^i)^2.  \label{gaufix}
\end{equation}
\noindent The free propagator of the gauge field is then given by

\begin{equation}
D_{\mu \nu }(k)=-\frac i\xi \frac{k_\mu k_\nu }{({\bf k}^2)^2}+\frac 1\Theta
\epsilon _{\mu \nu \rho }\frac{\overline{k}\,^\rho }{{\bf k}^2},
\label{propGFC}
\end{equation}
\noindent  where $\overline{k}=(0,k^1,k^2)$, which reduces, in the Landau
limit $(\xi \rightarrow \infty )$, to a totally antisymmetric form in the
Minkowski indices with the only nonvanishing components given by

\begin{equation}
D_{0i}(k)=-D_{i0}(k)=\frac 1\Theta \frac{\epsilon _{ij}k^j}{{\bf k}^2},
\label{propGFL}
\end{equation}
\noindent where $\epsilon _{ij}=\epsilon _{0ij}$. This gauge fixing
coincides with the choice of ref.\cite{bl} where the corresponding
nonrelativistic model is discussed. Notice also that, by requiring $\xi $ in
(\ref{gaufix}) to be dimensionless, we fix the mass dimension of $A_\mu $ as 
$1/2$ and so $\Theta $ as having dimension of mass. The other dimensional
parameters appearing in\ (\ref{lagran}) are $e$ and $\lambda $ which
dimensions are $1/2$ and $1$, respectively. In 2\ +\ 1 dimensions, the
bosonic matter field has mass dimension equal to $1/2$ and its free
propagator is given by $\Delta (p)=i\,(p^2-m^2+i\varepsilon )^{-1}. $
Besides the scalar field quadrilinear self-interaction vertex, minimal gauge
coupling with $A_\mu $ generates two kinds of vertices. The corresponding
Feynman rules are presented in Fig. 1, where we use a wavy line to represent
the gauge field propagator.

The form of the gauge field propagator (\ref{propGFL}) also interferes in
the power counting. For a generic graph, the degree of superficial
divergence is given by $d(G)=3-N_\phi /2-N_A-V_{\phi ^4}\,$, where $N_\phi $
and $N_A$ are the numbers of external particle and gauge field lines and $%
V_{\phi ^4}$ is the number of self-interaction vertices of the diagram. As
we will see, the 1-loop order self-energy corrections of both matter and
gauge fields are linearly divergent whereas the vertex correction is finite,
although it is superficially divergent. The particle four-point function has
also divergent contributions. In fact, the 1-loop graphs containing the
seagull vertex are linearly divergent. Therefore, the inclusion of the bare $%
\phi ^4$ vertex in the Lagrangian (\ref{lagran}) is necessary to render the
theory renormalizable.

Let us initially examine the 1-loop particle self-energy, the graphs of Fig.
2. Owning to the antisymmetry of the gauge field propagator (\ref{propGFL}),
the contributions that represent emission and absorption of a virtual gauge
particle are all identically null, a fact that does not happen in the
fermionic case \cite{hag,fle}. The tadpole with a gauge
propagator neck vanishes by charge conjugation so that the 1-loop particle
self-energy reduces to the tadpole graph and a simple ultraviolet cutoff $%
\Lambda _0$ introduced in the $|{\bf k}|$ integration gives

\begin{equation}
\widehat{\Sigma }^{(1)}(m;\Lambda _0)=\frac{-i\lambda }2\int^{\Lambda _0}%
\frac{d^3k}{(2\pi )^3}\,\frac i{[k^2-m^2+i\epsilon ]}
=\frac{i\lambda }{8\pi }m-\frac{i\lambda }{8\pi }\Lambda _0.  \label{SEcutoff}
\end{equation}

\noindent Had we used dimensional regularization, which acts as a
renormalization at one loop level in odd dimensions, we would have obtained
only the first term of the above expression. We see that $\widehat{\Sigma }%
^{(1)}$ does not depend on the external momentum and thus there is no wave
function renormalization at one loop level. The mass counterterm can be
written as $-\delta m^2\phi ^{*}\phi $ with $\delta m^2=-i\widehat{\Sigma }%
^{(1)}$ .

The one loop self-energy of the gauge field, the vacuum polarization, has
two linearly divergent parcels represented by the diagrams shown in Fig. 3.
It does not involve the gauge field propagator and so it is independent of
the gauge fixing chosen. Dimensional renormalization generates, naturally, a
gauge independent result which is given by

\begin{eqnarray}
\Pi _{\mu \nu }^{(1)}(q) &=&e^2\int \frac{d^3k}{(2\pi )^3}\frac{(2k+q)_\mu
\,(2k+q)_\nu }{\left[ k^2-m^2+i\epsilon \right] \,\left[
(k+q)^2-m^2+i\epsilon )\right] }  \nonumber \\
&&\hspace{1.0in}-2e^2\int \frac{d^3k}{(2\pi )^3}\frac{g_{\mu \nu }}{\left[
k^2-m^2+i\epsilon \right] } \nonumber \\
&=&-\frac{ie^2}{2\pi }m\left( 1-\int_0^1dx\sqrt{%
1-q^2x(1-x)/m^2}\right) \frac 1{q^2}\left[ q^2g_{\mu \nu }-q_\mu q_\nu
\right] .  \label{vacpol}
\end{eqnarray}

\noindent In the low momentum regime, $\left| q^2\right| \ll m^2$ , one has

\begin{equation}
\Pi _{\mu \nu }^{(1)}(q)\simeq -\frac{ie^2}{24\pi m}\left( 1+\frac{q^2}{20m^2%
}\right) \left[ q^2g_{\mu \nu }-q_\mu q_\nu \right] .  \label{vacpolLQ}
\end{equation}
If a gauge non invariant cutoff regularization is employed, an additional
linear divergent contribution ($\frac{ie^2}{6\pi }\Lambda _0g_{\mu \nu }$)
arises and the transversality of the vacuum polarization is lost. It should
be noticed that, the transverse nature of the finite part (\ref{vacpol})
requires the inclusion of the tadpole diagram. For this reason, we found
more convenient to consider the ordinary product of the operators in the
interaction Lagrangian instead of the Wick ordering. One should also notice
that, only a Maxwell term is generated by radiative corrections to the gauge
field propagator, in contrast with the fermionic case where a Chern-Simons
term is produced too \cite{djt,red}.

The 1-loop correction to the vectorial gauge coupling vertex, shown in Fig.
4, is finite. The other possible 1-loop contribution, which has one
trilinear and one self-interaction vertex inserted in a particle loop, is
null by charge conjugation. The sum of the two first parcels, the
contribution involving the seagull vertex, 

\begin{equation}
\Gamma _{3(S)}^{(1)\,\mu }=\frac{2ie^3}\Theta \int \frac{d^3k}{(2\pi )^3}%
\frac{(2p-k)^\sigma \,\epsilon _{\sigma \nu \rho }\,\overline{k}\,^\rho
\,g^{\nu \mu }}{{\bf k}^2\,\left[ (p-k)^2-m^2+i\epsilon \right] }-\left[
p\leftrightarrow p^{\prime }\right] \; ,  \label{gamaS}
\end{equation}

\noindent can be exactly calculated and, for external particles legs
in the mass shell, is given by $\Gamma _{3(S)}^{(1)0}=0$ and

\begin{equation}
\Gamma _{3(S)}^{(1)l}=\frac{-e^3}{2\pi \Theta }\epsilon _{ij}g^{jl}\left[ 
\frac{p^i\sqrt{m^2+{\bf p}^2}}{\sqrt{m^2}+\sqrt{m^2+{\bf p}^2}}-\frac{%
p^{\prime \,i}\sqrt{m^2+{\bf p}^{\prime \,2}}}{\sqrt{m^2}+\sqrt{m^2+{\bf p}%
^{\prime \,2}}}\right]  \label{gama3lS}
\end{equation}

\noindent for $l=1,2$. Although this contribution has not been considered in
the calculation of the anomalous magnetic moment using the covariant Landau
gauge in ref.\cite{sem}, it is relevant in the present case. The
triangle graph contribution, again for external momenta in the mass shell,
can be expressed in terms of Feynman integrals as

\begin{eqnarray}
\Gamma _{3(T)}^{(1)\,\mu }&=&\frac{ie^3}\Theta \int \frac{d^3k}{(2\pi )^3}%
\frac{(2p-k)^\sigma \,\epsilon _{\sigma \nu \rho }\,\overline{k}\,^\rho
\,(2p^{\prime }-k)^\nu \,(p+p^{\prime }-2k)^\mu }{{\bf k}^2\,\left[
(p-k)^2-m^2+i\epsilon \right] \,\left[ (p^{\prime }-k)^2-m^2+i\epsilon
)\right] } \nonumber \\
&=&\frac{e^3}{8\pi \Theta }\int_0^1dx\int_0^1dy\left[ 
\frac{A^\mu }{\left( Q_0^2-y{\bf Q}^2\right) ^{3/2}}+\frac{B^\mu }{\left(
Q_0^2-y{\bf Q}^2\right) ^{1/2}}\right]  \label{gama3T}
\end{eqnarray}
with the numerators given by 
\begin{eqnarray}
A^0 &=&\epsilon _{ij}p^iq^jQ^0\left( P^0-2Q^0\right) \hspace{0.3cm},
\label{A0} \\
B^0 &=&-2\epsilon _{ij}p^iq^j\hspace{0.3cm},  \label{B0} \\
A^l &=&\epsilon _{ij}p^iq^jQ^0\left( P^l-2yQ^l\right) \hspace{0.3cm},
\label{Al} \\
B^l &=&\epsilon _{ij}\left( 4V^i+2Q^0q^i\right) g^{jl}\hspace{0.3cm},
\label{Bl}
\end{eqnarray}
where $q^\mu =p^\mu -p^{\prime \mu }$, $P^\mu =p^\mu +p^{\prime \mu }$, $%
V^i=q^0p^i-p^0q^i$ and $Q^\mu (x)=p^\mu -q^\mu \left( 1-x\right) $. The $y$
integration in (\ref{gama3T}) can be easily done but the remaining $x$
integration is very complicated. We do not push any harder to get an exact
result, but rather, we shall look directly for the small momenta behavior.
For sake of simplicity, we restrict ourselves to the situation with $%
p^0=p^{\prime \,0}$ which implies ${\bf p}^2={\bf p}^{\prime \,2}$ for
particles in the mass shell. Actually, this is the situation we shall find
when dealing with the vertex insertions into the tree level
particle-particle scattering in the center of mass frame (Sec. 4). In this
case, $Q_0^2=m^2+{\bf p}^2$ and expanding (\ref{gama3lS}) and (\ref{gama3T})
for $|{\bf p}|/m$ small one obtains the total trilinear vertex correction,
up to order ${\bf p}^2/m^2$, as 
\begin{equation}
\left. \Gamma _3^{(1)0}(p,p-q)\right| _{q^0=0}\simeq \frac{-e^3}{4\pi \Theta 
}\left[ \frac{\epsilon _{ij}p^iq^j}m\right]  \label{gama30}
\end{equation}
and 
\begin{eqnarray}
\left. \Gamma _3^{(1)l}(p,p-q)\right| _{q^0=0} &\simeq &\frac{-e^3}{4\pi
\Theta }\left[ \epsilon _{ij}q^ig^{jl}\right] \left( 2+\frac 1{12}(5+\cos
\theta )\frac{{\bf p}^2}{m^2}\right)  \nonumber \\
&&+\frac{e^3}{4\pi \Theta }\left[ \frac{\epsilon _{ij}p^iq^j}m\right] \frac{%
(p+p^{\prime })^l}{4m}\hspace{0.3cm},  \label{gama3l}
\end{eqnarray}
where $\theta $ is the angle between the vectors ${\bf p}$ and ${\bf p}%
^{\prime }$.

The one loop correction to the seagull vertex, presented in Fig. 5, is of
fourth order in the charge $e$ and so it does not participate in the
particle-particle scattering at one loop level. No essential new
contribution is expected from this correction since, by gauge invariance, it
must be in accordance with the trilinear vertex one. In fact, by noticing
that only the second parcel in Fig. 2 has the external momentum flowing
through the diagram, one can readily verify the Ward identities at one loop
level

\begin{equation}
-\frac 1e\Gamma _{3\mu }^{(1)}(p,p)=\frac{\partial \widehat{\Sigma }^{(1)}}{%
\partial p^\mu }\hspace{0.3cm},  \label{W1}
\end{equation}

\begin{equation}
-\frac 1{e^2}\Gamma _{4\mu \nu }^{(1)}(p,p;0)=\frac{\partial ^2\widehat{%
\Sigma }^{(1)}}{\partial p^\mu \partial p^\nu }\hspace{0.3cm}.  \label{W2}
\end{equation}
Owing to the particular gauge we fixed, the above quantities are identically
null. Notice also that, there exist corrections to the quadrilinear gauge
vertex with one self-interaction vertex and one seagull or two trilinear
vertices inserted into a particle loop which do not vanish separately due to
charge conjugation. However, because of the Ward relation

\begin{equation}
e\frac \partial {\partial q_\nu }\Gamma _3^{(1)\mu }=\left. \Gamma
_4^{(1)\mu \nu }\right| _{q^{\prime }=q}\hspace{0.4cm},  \label{W3}
\end{equation}
their sum does not contribute to the vertex correction.

There remains to discuss the correction to the self-interaction vertex which
represents the 2-particle scattering in one loop order. This four-point
function, which will be calculated in Sec. IV, is linearly divergent and
therefore a counterterm of the form $C(\Lambda _0)\,(\phi ^{*}\phi )^2$ has
to be introduced in the Lagrangian. Counterterms linear in $p^\mu $ , the
first order Taylor subtraction in the BPHZ scheme, are absent due to the
rotational invariance, a fact that can be explicitly verified. The
particle-particle scattering, for low momenta, will be calculated using the
approximation scheme we shall now describe.

\section{Nonrelativistic Approximation}

The nonrelativistic approximation scheme we shall employ to calculate
particle-particle amplitudes in one loop order consists in the following
steps \cite{go}. First of all, denoting by $k$ the loop momentum, we
integrate over $k^0$ without making any restriction in order to guarantee
locality in time. This integration is greatly facilitated in the gauge we
are working since $k^0$ appears in the integrand of the amplitudes only via
the particle propagator and the trilinear vertex factors. The advantage in
using Coulomb gauge to discuss the nonrelativistic limit is that it gives
naturally a static interaction between the particles.

The remaining integration over the Euclidean ${\bf k}$ plane is then
separated into two parcels through the introduction of an intermediate
cutoff $\Lambda $ in the $|{\bf k}|$ integration. This auxiliary cutoff $%
\Lambda $ is chosen such that it satisfies the conditions

\begin{equation}
(i)\hspace{0.5cm}|{\bf p}|\ll \Lambda \ll m\hspace{0.5cm}\text{and}%
\hspace{0.5cm}(ii)\hspace{0.5cm}\left( \frac{|{\bf p}|}\Lambda \right)
^2\approx \left( \frac \Lambda m\right) ^2\approx \frac{|{\bf p}|}m\approx
\eta \hspace{0.3cm},  \label{cutoff}
\end{equation}
where $|{\bf p}|$ stands for the external nonrelativistic momenta, $m$ is
the renormalized particle mass and $\eta $ is established as the small
expansion parameter. These conditions can be easily met in nonrelativistic
condensed matter systems \cite{ct}. Clearly, the cutoff $\Lambda $
splits the space of the intermediate states into two parts. The circle of
radius $\Lambda $ and center at the origin of the ${\bf k}$-plane will be
referred as the low (L) energy sector whereas the $|{\bf k}|$ $>\Lambda $
region is the high (H) energy one.

The approximation procedure we work with rely heavily on the intrinsic
nature of each of these two sectors. In the low energy one, where all the
spatial momenta involved are small ($|{\bf p}|/m$ $,$ $|{\bf k}|/m\ll 1$),
we perform a $1/m$ expansion of the integrand. This means, for example, that
the free particle energy dispersion relation can be expanded, in the L-
sector, as

\begin{equation}
w_k=\sqrt{m^2+{\bf k}^2}=m+\frac{{\bf k}^2}{2m}-\frac{({\bf k}^2)^2}{8m^3}%
+...  \label{expanwk}
\end{equation}
\noindent The expanded integrand is then integrated, term by term, in the
region $0<|{\bf k}|<\Lambda $ and, naturally, a $\Lambda $- dependent result
is obtained. A further expansion in $\Lambda /m$ may be necessary to get the 
$\eta $ expansion of the L- contribution to the amplitude up to the desired
order.

In the H- region, $|{\bf k}|\gg |{\bf p}|$ and the integrand can be expanded
in a Taylor series around $|{\bf p}|=0$. This permits analytical
calculations of the integrals in every order in $\eta $. Certainly, this
expansion does not suppress ultraviolet divergences so, in that case, a
regularization procedure is assumed. The result is also $\Lambda $-
dependent and one must expand in $\Lambda /m$ to obtain the $\eta $
expansion of the H- contribution. However, as we should expect for sake of
consistence of the nonrelativistic approximation, the $\Lambda $- dependent
parcels of the L and the H contributions of each diagram cancel identically.
Apparently the two sectors L and H are treated differently but, in fact, the
same $1/m$ expansion is made and the exact cancellation of the $\Lambda $-
dependent terms reflects just the additivity property of the integration.
The complete amplitude obtained by adding the contributions of the L and H
sectors is the correct $|{\bf p}|/m$ expansion up to the order we have
worked.

It should be noticed that by choosing a distinct routing of the external
momenta through the diagram (or by using the Feynman parametrization to
simplify the integrand before making the $k^0$ integration) one gets
different L and H contributions for the graph. The changes, however, occur
only in the coefficients of the $(|{\bf p}|/\Lambda )^n$ which are not
relevant for the reduction process we shall discuss later.

We can apply this scheme to the calculation of the one loop self-energy,
vacuum polarization and vertex corrections and identify the origin of each
of them. For the self-energy, the separation of the L- and H- contributions
gives, up to the order $\eta ^2\approx {\bf p}^2/m^2$,

\begin{equation}
\widehat{\Sigma }_L^{(1)}(m;\Lambda )\simeq \frac{i\lambda }{16\pi }m\left[ -%
\frac{\Lambda ^2}{m^2}+\frac 14\frac{\Lambda ^4}{m^4}\right]  \label{SElow}
\end{equation}
\noindent and

\begin{equation}
\widehat{\Sigma }_H^{(1)}(m;\Lambda ,\Lambda _0)\simeq \frac{i\lambda }{%
16\pi }m\left[ \frac{\Lambda ^2}{m^2}-\frac 14\frac{\Lambda ^4}{m^4}\right] +%
\frac{i\lambda }{8\pi }m-\frac{i\lambda }{8\pi }\Lambda _0\hspace{0.3cm}.
\label{SEhigh}
\end{equation}
We see that the low energy intermediate states contribution is polynomial in 
$\Lambda /m$ and that both finite and divergent (as $\Lambda _0\rightarrow
\infty $) parts of (\ref{SEcutoff}) come from the high energy sector. This
also happens with the vacuum polarization, for which one has (in the case of 
$|q^2|\ll m^2$)

\begin{eqnarray}
\Pi _L^{(1)\mu \nu }(q;\Lambda ) &\simeq &\frac{ie^2}{6\pi }mg^{\mu \nu
}\left[ \frac{\Lambda ^2}{m^2}-\frac 12\frac{\Lambda ^4}{m^4}-\frac 14\frac{%
q^2}{m^2}\left( \frac 13\frac{\Lambda ^2}{m^2}-\frac 14\frac{\Lambda ^4}{m^4}%
\right) \right]  \nonumber \\
&&+\frac{ie^2}{16\pi }\frac{q^\mu q^\nu }m\left[ \frac 13\frac{\Lambda ^2}{%
m^2}-\frac 14\frac{\Lambda ^4}{m^4}\right]  \label{VPlow}
\end{eqnarray}

\noindent and\ 
\begin{eqnarray}
\Pi _H^{(1)\mu \nu }(q;\Lambda ,\Lambda _0) &\simeq &-\frac{ie^2}{24\pi m}%
\left( 1+\frac{q^2}{20m^2}\right) \left[ q^2g^{\mu \nu }-q^\mu q^\nu \right]
\nonumber \\
&&\ +\frac{ie^2}{6\pi }\Lambda _0g^{\mu \nu }-\Pi _L^{(1)\mu \nu }(q;\Lambda
)\hspace{0.3cm}.  \label{VPhigh}
\end{eqnarray}
The L- contribution to the trilinear vertex correction is given by

\begin{equation}
\Gamma _{3L}^{(1)0}\simeq 0\hspace{0.6cm}\text{and}\hspace{0.6cm}\Gamma
_{3L}^{(1)l}\simeq \frac{-e^3}{4\pi \Theta }\left[ \epsilon
_{ij}q^ig^{jl}\right] \left\{ \frac{\Lambda ^2}{m^2}-\frac 34\frac{\Lambda ^4%
}{m^4}\right\} \hspace{0.3cm},  \label{Vlow}
\end{equation}
and again the whole correction, equations (\ref{gama30}) and (\ref{gama3l}),
comes from the high energy intermediate states. In the above equations, and
from now on, the symbol $\simeq $ denotes that the expression which follows
holds up to the order $\eta ^2$.

The fact that the low energy intermediate states contributions to the basic
radiative corrections are small and suppressed by part of the high ones
reflects the nonrelativistic nature of the L sector. The corresponding
nonrelativistic field theory, that is the theory with a Galilean invariant,
Schr\"odinger, kinematics and the same interactions, has a distinct
intrinsic nature. Particle propagation is only forward in time and
interactions are instantaneous. This means that there are no dynamical
corrections to the mass and to the charge of the particles which are given
phenomenological parameters of the theory.

There is no objection, in principle, to extend the procedure described above
to obtain the nonrelativistic approximation, that is the $|{\bf p}|/m$
expansion, of quantum amplitudes in any order of perturbation theory. One
naturally needs to introduce an intermediate cutoff for each independent
loop integration and, certainly, makes use of the Feynman parametrization to
symmetrize the integrand in order to perform the $k^0$ integration . An
example can be found in ref. \cite{go}, where the two loop self-energy
of the $\phi ^4$ theory was calculated.

\section{Particle-Particle Scattering}

To simplify calculations, we choose to work in the center of mass (CM)\
frame with external particles on the mass shell, that is ${\bf p}_1=-{\bf p}%
_2={\bf p}$ , ${\bf p}_1^{\prime }=-{\bf p}_2^{\prime }={\bf p}^{\prime }$
and $p_1^0=p_2^0=p_1^{\prime \,0}=p_2^{\prime \,0}=w_p=\sqrt{m^2+{\bf p}^2%
\text{ }}$ . The tree level particle-particle amplitude has three
contributions, presented in Fig. 6, and is given by

\begin{eqnarray}
A^{(0)} &=&-\lambda +\left\{ \left. i\,e^2\,(p_1+p_1^{\prime })^\mu \,D_{\mu
\nu }(p_1-p_1^{\prime })\,(p_2+p_2^{\prime })^\nu \right| _{{\rm CM}}+\left.
\left[ p_1^{\prime }\leftrightarrow p_2^{\prime }\right] \right| _{{\rm CM}%
}\right\} \;  \nonumber  \\
\ &=&-\lambda -i\frac{8e^2}\Theta \sqrt{m^2+{\bf p}^2}\cot \theta \simeq
-\lambda -i\frac{8e^2}\Theta m\left( 1+\frac{{\bf p}^2}{2m^2}\right) \cot
\theta \;,  \label{Atree}
\end{eqnarray}

\noindent where $\theta $ is the scattering angle and $m$ is the
renormalized mass of the bosonic particle. By definiteness, we take the
amplitude as being $(-i)$ times the 1PI four point function. This choice is
only to facilitate the comparison with the nonrelativistic case discussed in
ref.\cite{bl}.

We shall now calculate the 1-loop order scattering amplitude, for low external
momenta and up to order $\eta ^2\approx {\bf p}^2/m^2$. Consider firstly the
amplitude that comes from the minimally subtracted vacuum
polarization and gauge vertex corrections inserted into the tree level
diagrams. Using equation (\ref{vacpolLQ}) for the polarization tensor with $%
q^0=0$, one immediately gets the contribution to the amplitude due to the
one loop correction to the gauge field propagator as

\begin{equation}
A^{(p)}\simeq -\frac{2e^4}{\pi \Theta ^2}m\left\{ \frac 16+\frac 7{30}\frac{%
{\bf p}^2}{m^2}\right\} \hspace{0.4cm}.  \label{ApT}
\end{equation}
Similarly, the trilinear vertex correction $\Lambda _3^{(1)\mu }$, given by (%
\ref{gama30}) and (\ref{gama3l}), leads to the following contribution to the
one loop amplitude

\begin{equation}
A^{(v)}\simeq -\frac{2e^4}{\pi \Theta ^2}m\left\{ 2+\frac 53
\frac{{\bf p}^2}{m^2}\right\} \hspace{0.4cm}.  \label{AvT}
\end{equation}

Diagrams, like those shown in Fig. 7, that admixes particle self-interaction
and gauge field exchange, do not contribute. The first vanishes by charge
conjugation, the second is null due to the antisymmetric form of the gauge
field propagator whereas the chalice diagram gives 

\begin{equation}
A^{(ch)}=-i\frac{e^2 \lambda}{8\pi \Theta} \sin \theta 
\frac{{\bf p}^2}{m^2} + 
(\theta \rightarrow \pi + \theta )= 0 \,\,\, .
\end{equation}
This contribution, and also all the terms proportional to $\sin
\theta$ (or $\cos \theta$) appearing in the other graphs, vanishes due
to due to the required symmetrization in the outgoing particles. The non
admixture of the gauge coupling vertices and the self-interaction one (at
least in one loop level) means, as we are going to see, that $\lambda \phi
^4 $ interaction comes about only to renormalize the particle-particle
scattering due to CS gauge field exchange.

The most important one loop particle-particle scattering comes from the
diagrams shown in Fig. 8 and, to calculate these contributions to the
2-particle amplitude, we shall use the approximation scheme described in the
last section. The group $(a)$ is the finite self-interaction scattering
which was discussed before \cite{go}. Adding separately the L- and the
H- contributions of each graph, one obtains

\begin{equation}
A_L^{(a)}\simeq \frac{\lambda ^2}{32\pi m}\left\{ \left( 1-\frac{{\bf p}^2}{%
2m^2}\right) \left[ \ln \left( \frac{\Lambda ^2}{{\bf p}^2}\right) +i\pi
\right] -\frac{{\bf p}^2}{\Lambda ^2}-\frac{{\bf p}^4}{2\Lambda ^4}+\frac{%
3\Lambda ^2}{2m^2}-\frac{21\Lambda ^4}{16m^4}\right\} ,  \label{AaL}
\end{equation}

\begin{equation}
A_H^{(a)}\simeq \frac{\lambda ^2}{32\pi m}\left\{ -\left( 1-\frac{{\bf p}^2}{%
2m^2}\right) \ln \left( \frac{\Lambda ^2}{4m^2}\right) +\frac{{\bf p}^2}{%
\Lambda ^2}+\frac{{\bf p}^4}{2\Lambda ^4}-\frac{3\Lambda ^2}{2m^2}+\frac{%
21\Lambda ^4}{16m^4}+4-\frac{{\bf p}^2}{6m^2}\right\}  \label{AaH}
\end{equation}
\noindent and, thus, $A^{(a)}=A_L^{(a)}+A_H^{(a)}$ is given by

\begin{equation}
A^{(a)}\simeq \frac{\lambda ^2}{32\pi m}\left\{ \left( 1-\frac{{\bf p}^2}{%
2m^2}\right) \left[ \ln \left( \frac{4m^2}{{\bf p}^2}\right) +i\pi \right]
+4-\frac{{\bf p}^2}{6m^2}\right\} .  \label{Aa}
\end{equation}

\noindent As we have stressed in ref.\cite{go}, the intermediary cutoff
procedure generates the correct $|{\bf p}|/m$ expansion of the amplitudes in
the case of $\lambda \phi ^4$ theory alone. In the present case, and
certainly in many other theories, where analytical results are very
difficult (if not impossible) to be achieved, our procedure emerges as an
useful calculational tool to generate the $|{\bf p}|/m$ expansion of the
amplitudes, having the exact result in the $\phi ^4$ case as a basis of
confidence.

The box diagrams of Fig. 8$(b)$ are also calculated separately. Let us first
concentrate in the ``right'' box diagram corresponding to the direct
exchange of two virtual gauge particles, the first parcel of Fig. 8$(b)$.
After performing the $k^0$ and the angular integrations, by using the
Cauchy-Gousart theorem and splitting the $|{\bf k}|$ integration into its
low and high energy contributions one is led, by adding the $({\bf p}%
^{\prime }\longleftrightarrow -{\bf p}^{\prime })$ companion, to

\begin{eqnarray}
A_L^{(b)dir} &\simeq &-\frac{2e^4}{\pi \Theta ^2}m\left\{ \left( 1+\frac{%
{\bf p}^2}{2m^2}\right) \left[ \ln (2|\sin \theta |)+i\pi \right] -\frac{%
{\bf p}^2}{m^2}\right.  \nonumber \\
&&\ \left. \hspace{1.5cm}-\frac 12\cos \theta \ln \left( \frac{1-\cos \theta 
}{1+\cos \theta }\right) \frac{{\bf p}^2}{m^2}-(1-2\cos ^2\theta )\frac{{\bf %
p}^4}{\Lambda ^4}\right\}  \label{AbdirL}
\end{eqnarray}

\noindent and

\begin{equation}
A_H^{(b)dir}\simeq -\frac{2e^4}{\pi \Theta ^2}m\left\{ (1-2\cos ^2\theta )%
\frac{{\bf p}^4}{\Lambda ^4}\right\} .  \label{AbdirH}
\end{equation}

\noindent Adding (\ref{AbdirL}) and (\ref{AbdirH}), we get

\begin{eqnarray}
A^{(b)dir} &\simeq &-\frac{2e^4}{\pi \Theta ^2}m\left\{ \left( 1+\frac{{\bf p%
}^2}{2m^2}\right) \left[ \ln (2|\sin \theta |)+i\pi \right] -\frac{{\bf p}^2%
}{m^2}\right.  \nonumber \\
&&\ \left. \hspace{1.5cm}-\frac 12\cos \theta \ln \left( \frac{1-\cos \theta 
}{1+\cos \theta }\right) \frac{{\bf p}^2}{m^2}\right\} .  \label{Abdir}
\end{eqnarray}

\noindent We see that the contribution to the amplitude coming from the
direct box diagram comes entirely from the low energy, nonrelativistic,
sector. The main steps of these calculations (and the others to come) are
outlined in the Appendix.

The contrary happens with the twisted box diagram, the second in Fig. 8$(b)$%
. The $k^0$ integration can be easily done as a contour integration but the
resulting ${\bf k}$ integration has a rather non trivial angular part. The
alternative possibility of introducing Feynman parameters allows the
evaluation of the $k$ integration but the parametric integrals that remain
are intractable. By approximating the integrands as described in the last
section, one simplifies the angular and the radial integrations that appear
obtaining, after adding its final particles exchanged partner,

\begin{equation}
A_L^{(b)twist}\simeq 0\hspace{0.6cm}\text{and}\hspace{0.6cm}%
A^{(b)twist}\simeq A_H^{(b)twist}\simeq -\frac{2e^4}{\pi \Theta ^2}m\left\{ 
\frac{{\bf p}^2}{2m^2}\right\} \hspace{0.3cm}.  \label{AbtwistLH}
\end{equation}

\noindent As in the case of the last two graphs of Fig. 8$(a)$, diagrams
that involve propagation backward in time possess a small, $\Lambda $
dependent, contribution coming from the L sector. For the twisted box
diagram this contribution is of order $\eta ^3.$ The total box amplitude, $%
A^{(b)}=A^{(b)dir}+A^{(b)twist}$, is finite and, up to order ${\bf p}^2/m^2$%
, is given by

\begin{eqnarray}
A^{(b)} &\simeq &-\frac{2e^4}{\pi \Theta ^2}m\left\{ \left( 1+\frac{{\bf p}^2%
}{2m^2}\right) \left[ \ln (2|\sin \theta |)+i\pi \right] -\frac{{\bf p}^2}{%
2m^2}\right.  \nonumber \\
&&\ \ \left. \hspace{1.5cm}-\frac 12\cos \theta \ln \left( \frac{1-\cos
\theta }{1+\cos \theta }\right) \frac{{\bf p}^2}{m^2}\right\} \hspace{0.3cm}.
\label{Abox}
\end{eqnarray}

The third group, the seagull scattering $(c)$, has to be treated more
carefully since it carries the divergence of the four point function. One
can immediately see that the $k^0$ integration of the first diagram, the
gauge bubble, would diverge if made alone. However, the two triangle
diagrams, which give identical contributions, have also divergent $k^0$
integrations that exactly compensates the divergence of the first graph.
Therefore, by taking all the diagrams of group $(c)$ together, there is no
need to maintain any cutoff and the $k^0$ integration is unrestricted. The
remaining ${\bf k}$ integration has two parts. One is finite and has the
same angular integration as that appearing in the direct box. The other one,
with a distinct but feasible angular integration, is infinite so that an
ultraviolet cutoff $\Lambda _0$ has to be introduced in the $|{\bf k}|$
integration of the H sector. Up to order $\eta ^2$, the low energy
intermediate states contribution for the seagull scattering is given by

\begin{eqnarray}
A_L^{(c)} &\simeq &-\frac{2e^4}{\pi \Theta ^2}m\left\{ \left( 1+\frac{{\bf p}%
^2}{2m^2}\right) \left[ \ln \left( \frac{\Lambda ^2}{{\bf p}^2}\right) -\ln
(2|\sin \theta |)\right] +\frac{{\bf p}^2}{m^2}\right.  \nonumber \\
&&\ \ \left. \hspace{1.5cm}+\frac 12\cos \theta \ln \left( \frac{1-\cos
\theta }{1+\cos \theta }\right) \frac{{\bf p}^2}{m^2}+\frac 12(1-2\cos
^2\theta )\frac{{\bf p}^4}{\Lambda ^4}+\frac{\Lambda ^4}{16m^4}\right\} .
\label{AcL}
\end{eqnarray}

\noindent The H sector contribution, cutoff regulated, is given by

\begin{eqnarray}
A_H^{(c)} &\simeq &-\frac{2e^4}{\pi \Theta ^2}m\left\{ -\left( 1+\frac{{\bf p%
}^2}{2m^2}\right) \ln \left( \frac{\Lambda ^2}{4m^2}\right) -1\right. 
\nonumber \\
&&\ \ \ \left. \hspace{1.5cm}-\frac 12(1-2\cos ^2\theta )\frac{{\bf p}^4}{%
\Lambda ^4}-\frac{\Lambda ^4}{16m^4}+\left[ \frac{\Lambda _0}m\right]
\right\}  \label{AcH}
\end{eqnarray}

\noindent and, thus, the total $(c)$ contribution to the amplitude is

\begin{eqnarray}
A^{(c)} &\simeq &-\frac{2e^4}{\pi \Theta ^2}m\left\{ \left( 1+\frac{{\bf p}^2%
}{2m^2}\right) \left[ \ln \left( \frac{4m^2}{{\bf p}^2}\right) -\ln (2|\sin
\theta |)\right] \right.  \nonumber \\
&&\ \ \ \ \left. \hspace{1.5cm}-1+\frac 12\cos \theta \ln \left( \frac{%
1-\cos \theta }{1+\cos \theta }\right) \frac{{\bf p}^2}{m^2}+\frac{{\bf p}^2%
}{m^2}\right\} -\frac{2e^4}{\pi \Theta ^2}\Lambda _0\;.  \label{Ac}
\end{eqnarray}

The constant divergent term above can be suppressed by a counterterm of the
form $\frac{e^4}{2\pi \Theta ^2}\Lambda _0(\phi ^{*}\phi )^2$ introduced in
the Lagrangian density. We can also imagine that the bare self-coupling $%
\lambda $ carries a divergent part that just cancel the divergence of the
four point function. In any case, we take the finite part of (\ref{Ac}) as
the 1-loop renormalized $(c)$ contribution. This would be the result if we
had used dimensional renormalization.

Before summing up all contributions to get the total one loop amplitude, let
us separate the total CS exchange scattering by adding the $(b)$ and $(c)$
parts. One gets 
\begin{equation}
A^{(CS)}\simeq -\frac{2e^4}{\pi \Theta ^2}m\left\{ \left( 1+\frac{{\bf p}^2}{%
2m^2}\right) \left[ \ln \left( \frac{4m^2}{{\bf p}^2}\right) +i\pi \right]
-1+\frac{{\bf p}^2}{2m^2}\right\} \hspace{0.3cm},  \label{ACS}
\end{equation}
and it is noticeable that the cancellation of the $\theta $ dependent terms
of the box and the seagull amplitudes happens in both dominant and
subdominant orders.

The total renormalized particle-particle scattering amplitude up to 1-loop, 
$A^{0}+A^{(a)}+A^{(CS)}+A^{(p)}+A^{(v)}$, up to order ${\bf p}^2/m^2$
is given by

\begin{eqnarray}
A^{(1)} &\simeq &
-\lambda -i\frac{8e^2}\Theta m\left( 1+\frac{{\bf p}^2}{2m^2}\right) \cot
\theta \nonumber \\
&&+m\left( \frac{\lambda ^2}{32\pi m^2}-\frac{2e^4}{\pi
\Theta ^2}\right) \left[ \ln \left( \frac{4m^2}{{\bf p}^2}\right) +i\pi
\right]  \nonumber \\
&&-m\left( \frac{\lambda ^2}{64\pi m^2}+\frac{e^4}{\pi \Theta ^2}\right) 
\frac{{\bf p}^2}{m^2}\left[ \ln \left( \frac{4m^2}{{\bf p}^2}\right) +i\pi
\right]  \nonumber \\
&&+m\left( \frac{\lambda ^2}{8\pi m^2}-\frac{7e^4}{3\pi \Theta ^2}\right)
-m\left( \frac{\lambda ^2}{192\pi m^2}+\frac{24e^4}{5\pi \Theta ^2}\right) 
\frac{{\bf p}^2}{m^2}\hspace{0.3cm}.  \label{Atotal}
\end{eqnarray}
The leading term of this expansion coincides with the result of ref. 
\cite{boz}. An important feature of the above result is that the
existence of critical values of the self-interaction parameter for which the
one loop scattering vanishes is restricted to the leading order. In other
words, there are critical values of $\lambda $, namely

\begin{equation}
\frac{\lambda _c^{\pm }}{4m^2}=\pm \frac{2e^2}{m|\Theta |}\hspace{0.3cm},
\label{ldc}
\end{equation}
for which the leading term of $A^{(1)}$ vanishes but the subdominant parcels
do not. By choosing $\lambda =\lambda _c^{+}$, the tree level amplitude
becomes the Aharonov-Bohm scattering for identical particles \cite{bl}.
The subdominant terms then represent relativistic corrections to the
Aharonov-Bohm scattering \cite{go2}. It should be pointed out that the
real constant term of (\ref{Atotal}) could be supressed by a finite
renormalization of the coupling constant $\lambda$.

\section{Nonrelativistic Reduction and the AB scattering}

The perturbative treatment of the AB scattering has long been known to be a
very delicate problem due to the singular nature of the potential \cite{fein}.
In the first quantized viewpoint \cite{qm},
perturbative renormalization requires the introduction of a delta function
potential which is equivalent to a $\phi ^4$ self-interaction in the field
theoretical approach \cite{bl,kile}. In the latter case
\cite{bl},
the nonrelativistic (NR) scattering amplitude, that is the 2-particle CM
scattering amplitude calculated with the Galilean invariant Lagrangian
density

\begin{equation}
{\cal L}_{NR}=\psi ^{*}\left( iD_t+\frac{{\bf D}^2}{2m}\right) \psi -\frac{%
v_0}4:(\psi ^{*}\psi )^2:+\frac \Theta 2\partial _t{\bf A}\times {\bf A}%
-\Theta A_0{\bf \nabla }\times {\bf A}  \label{lagranNR}
\end{equation}
and with the same gauge fixing (\ref{gaufix}) $(\xi \rightarrow \infty )$,
is given, at one loop level, by

\begin{equation}
{\cal A}_{NR}^{(1)}=\frac m{8\pi }\left( v_0^2-\frac{4e^4}{m^2\Theta ^2}%
\right) \left[ \ln \left( \frac{\Lambda _{NR}^2}{{\bf p}^2}\right) +i\pi
\right] \hspace{0.3cm},  \label{A1NRcof}
\end{equation}
where $\Lambda _{NR}$ is a nonrelativistic ultraviolet cutoff. The graphs
that enter in this calculation are the fish diagram (the first of Fig.
8(a)), the direct box and the two triangle diagrams with, in all of them,
the particle lines representing the nonrelativistic propagator

\begin{equation}
\Delta _{NR}(\omega ,{\bf k})=\frac i{\omega -{\bf k}^2/2m+i\varepsilon }\;.
\label{propNR}
\end{equation}
The renormalization is implemented by redefining the nonrelativistic
self-coupling constant, $v_0=v+\delta v$, so that the total renormalized
nonrelativistic amplitude, obtained by adding the tree level scattering, is
given, up to order $e^4$, by

\begin{equation}
{\cal A}_{NR}=-v-i\frac{2e^2}{m\Theta }\cot \theta +\frac m{8\pi }\left( v^2-%
\frac{4e^4}{m^2\Theta ^2}\right) \left[ \ln \left( \frac{\mu ^2}{{\bf p}^2}%
\right) +i\pi \right] \hspace{0.3cm},  \label{ANRren}
\end{equation}
where $\mu $ is an arbitrary mass scale, introduced by the renormalization,
that breaks the scale invariance of the amplitude \cite{bl}.

Prior to examine the similarities between the results of the relativistic
and the nonrelativistic models, the normalization of states has to be
properly adjusted. In the relativistic case one takes $\left\langle {\bf p}%
^{\prime }|{\bf p}\right\rangle =2w_p\delta ({\bf p}^{\prime }-{\bf p})$
while the usual normalization in a nonrelativistic theory does not have the $%
2w_p$ factor. Besides that, the cross-sections involve the relative velocity
and thus, for the purpose of comparison, the CM amplitudes calculated in the
last section must be multiplied by 
\begin{equation}
f\,\left( \frac 1{\sqrt{2w_p}}\right) ^4=\frac 1{4m^2}\left[ 1-\frac{3{\bf p}%
^2}{4m^2}+...\right] \hspace{0.3cm},  \label{norfac}
\end{equation}
where $f^2={w_p/m}$ is the ratio between the nonrelativistic and
relativistic velocities.

Let us now compare the renormalized nonrelativistic amplitude (\ref{ANRren})
with the leading term of the $|{\bf p}|/m$ expansion of the minimally
subtracted relativistic scattering (\ref{Atotal}) given by

\begin{equation}
{\cal A}^{{\rm lead}}=-\frac \lambda {4m^2}-i\frac{2e^2}{m\Theta }\cot
\theta +\frac m{8\pi }\left( \frac{\lambda ^2}{16m^4}-\frac{4e^4}{m^2\Theta
^2}\right) \left[ \ln \left( \frac{4m^2}{{\bf p}^2}\right) +i\pi \right] \;,
\label{Alead}
\end{equation}
where the calligraphic ${\cal A}$ means that the amplitude already
incorporates the nonrelativistic normalization factor (\ref{norfac}).

Confronting the tree levels, one sees that the self-interaction parameters
are related by

\begin{equation}
v=\frac \lambda {4m^2}\hspace{0.3cm}.  \label{vlamb}
\end{equation}
It is also immediately seen that, by fixing the nonrelativistic
renormalization point such that $\mu ^2=4m^2$, the two expressions coincide,
but this choice is completely arbitrary. What in fact coincides, when
comparing the 1-loop NR scattering with the leading order (in $|{\bf p}|/m$)
of the relativistic result, are the critical values for which the one loop
scattering vanishes, that is $v_c^{\pm }=\lambda _c^{\pm }/4m^2 
={\pm}2e^2/m|\Theta |$. For these
values of the self-coupling, which can be reached by a finite
renormalization, one regains the scale invariance of the NR amplitude and by
choosing the $v_c^{+}$ value, corresponding to a repulsive
contact interaction, the amplitude reduces to the Aharonov-Bohm amplitude
for identical particles which is given by \cite{bl} 
\begin{equation}
{\cal A}_{{\rm AB}}=-i\frac{4\pi }m\alpha \left[ \cot \theta -i{\rm sgn}%
(\alpha )\right] +{\cal O}\left( \alpha ^3\right) \;,  \label{AB}
\end{equation}
where $\alpha =e^2/2\pi \Theta $ is the Aharonov-Bohm parameter.

The subdominant terms that survive after fixing $\lambda =\lambda _c^{+}$
can be seen as relativistic field theoretical corrections to the
Aharonov-Bohm scattering \cite{go2} and are given by

\begin{eqnarray}
{\cal A}^{{\rm sub}} &=&\frac{\pi }m\,\alpha \,\left[ 3{\rm sgn\,}(\alpha )+
i \cot \theta \right] \frac{{\bf p}^2}{m^2} -\frac{2\pi }m\,\alpha ^2 \frac{%
{\bf p}^2}{m^2}\left[ \ln \left( \frac{ 4m^2}{{\bf p}^2}\right) +i\pi \right]
\nonumber \\
&&\ \ +\frac{17}3\frac \pi m \,\alpha ^2 -\frac{563}{60}\frac \pi m\,\alpha
^2\frac{{\bf p}^2}{m^2}\;.  \label{Asubled}
\end{eqnarray}
Part of the correction of the tree level $(\,\sim \alpha )$ is due to the
normalization of states and the relative velocity factor and so has a pure
kinematical origin, but not all of it since the scattering amplitude
corresponding to the exchange of one virtual gauge particle
(\ref{Atree}) depends on the
CM energy as a consequence of the minimal coupling. The other corrections
come from the 1-loop $(e^4)$ contribution to the perturbative
expansion. Terms proportional to $\alpha ^2$ do not exists in
nonrelativistic AB scattering (which exact result is function of $\sin \pi
\alpha \,$) and may be detected in experiments with fast particles.

Outside the critical values, the different ultraviolet structures of the
relativistic and the nonrelativistic models fully manifest. The
nonrelativistic triangle graph is logarithmically divergent while the
relativistic one is linearly divergent. The nonrelativistic fish diagram is
also logarithmically divergent whereas the relativistic (channel $s$) graph
is finite \cite{f1}. The distinct nature of the
divergences in the NR theory is due to its own kinematics reflected
in the form of the NR propagator (\ref{propNR}). The $1/m$
expansion we have used to calculate the low energy intermediate states
contributions mimics this aspect and one naturally
expects that the L-sector amplitudes could be mapped into a nonrelativistic
framework. Such a reduction process can be implemented and, certainly,
requires a reinterpretation of the intermediate cutoff $\Lambda $.

Let us initially concentrate on the leading order in $|{\bf p}|/m$
which should reproduce the NR case. As we saw, contributions coming
from the radiative corrections to propagators and vertices are
subdominant, come entirely from H sector (the L part is polynomial in 
$\Lambda ^2/m^2\approx \eta $) and so are neglected in agreement with
what is expected in the nonrelativistic theory, where the parameters
$m$, $e$ and $\Theta $ are fixed at their phenomenological
values. Now, by comparing the L-sector, leading order, contribution to
the 1-loop scattering,
 
\begin{equation}
{\cal A}_L^{(1)\,{\rm lead}}=\frac m{8\pi }\left( \frac{\lambda ^2}{16m^4}- 
\frac{4e^4}{m^2\Theta ^2}\right) \left[ \ln \left( \frac{\Lambda ^2}{{\bf p}
^2}\right) +i\pi \right] \hspace{0.3cm}.  \label{AscattL}
\end{equation}
with the NR amplitude (\ref{A1NRcof}), taking into account
(\ref{vlamb}), one sees that they coincide if the intermediate cutoff is
reinterpreted as a genuine nonrelativistic ultraviolet cutoff. In such
a mapping the L-sector contribution is identified with the NR
scattering amplitude while the H part can be seen as the needed
counterterm to render the nonrelativistic theory finite, renormalized.
In this way, the renormalization
of the nonrelativistic model of ref. \cite{bl} can be better understood.

This identification can be extended to a more general context by considering
a $d$-dimensional space--time. For example, the contribution of the L sector
to the $s$ channel amplitude (the first graph of Fig. 8(a)) and the NR fish
diagram are given by

\begin{eqnarray}
A_L^{(s)} &=&\frac{\lambda ^2\,m^{-1}}{2^{d+2}\,\pi ^{(d-1)/2}}\,\frac 1{%
\Gamma \left( \frac{d-1}2\right) }\,  \nonumber \\
&&\ \ \;\times \int_0^{\Lambda ^2}d({\bf k}^2)\,\frac{({\bf k}^2)^{(d-3)/2}}{%
{\bf k}^2-{\bf p}^2-i\epsilon }\,\left[ 1-\frac{{\bf k}^2}{2m^2}+\frac{3{\bf %
k}^4}{8m^4}-\cdots \right]  \label{AsLdimD}
\end{eqnarray}
and

\begin{equation}
{\cal A}_{{\rm NR}}^{fish}=\frac{v_0^2\,m}{2^d\,\pi ^{(d-1)/2}}\,\frac 1{%
\Gamma \left( \frac{d-1}2\right) }\,\int_0^{\Lambda _{{\rm NR}}^2}d({\bf k}%
^2)\,\frac{({\bf k}^2)^{(d-3)/2}}{{\bf k}^2-{\bf p}^2-i\epsilon }\,\;.
\label{Apeixe}
\end{equation}
For $d=1+1$, the nonrelativistic scattering is finite and coincides with the
leading term of the $\eta $ expansion (in fact a $\eta ^{1/2}$ expansion for 
$d=1+1$) of the L contribution to the $s$ channel. In this case, rigorous
results for the $\phi ^4$ theory can be found in ref. \cite{dim}. For $%
d>2+1$, one sees that the identification $\Lambda \leftrightarrow \Lambda _{%
{\rm {NR}}}$ can also be done and produces the same expressions. This
mapping can figuratively be pictured as a NR lens which magnifies $\Lambda $
to $\Lambda _{{\rm {NR}}}$ ($\rightarrow \infty $) making kinematics be
nonrelativistic. From a pragmatic viewpoint, one sees that the
nonrelativistic limit can be obtained by calculating the L contribution to
the scattering amplitude using the cutoff procedure described in Sec. III. \ 

We shall now investigate the possibility of constructing an effective
nonrelativistic Lagrangian that reproduces the scattering amplitudes
incorporating the relativistic corrections up to order ${\bf p}^2/m^2$ . To
account for the next-to-leading contribution to the amplitudes in the
nonrelativistic context one has to add counterterms and new interactions to
the usual nonrelativistic Lagrangian (\ref{lagranNR}). As a general
strategy, we consider the H-sector contributions to the relativistic
scattering amplitudes as counterterms in the nonrelativistic theory and
search for Galilean invariant effective interactions, preserving gauge
invariance, to generate nonrelativistic scattering processes which compare
with the L-sector amplitude. From beginning, we expect these terms to be
nonrenormalizable interactions, like happen in effective field theories 
\cite{lep}.

The subdominant, nonpolynomial, part of the L-sector contribution to the
two-body scattering is given, neglecting terms that are powers of $\eta $
involving $\Lambda $, by

\begin{equation}
{\cal A}_L^{(1)\,{\rm subdom}}\simeq \ -\frac m{8\pi }\left( \frac{5\lambda
^2}{64m^4}-\frac{e^4}{m^2\Theta ^2}\right) \frac{{\bf p}^2}{m^2}\left[ \ln
\left( \frac{\Lambda ^2}{{\bf p}^2}\right) +i\pi \right] \;.
\label{ALsubdom}
\end{equation}
Notice that the subdominant polynomial part is cutoff-dependent and
this freedom can be used to eliminate it.  The term inside the square
bracket in this expression appears in both fish
and triangle NR diagrams. This suggests, in order to account for the ${\bf p}%
^2$ factor, that the effective interactions which might reproduce these one
loop contributions should contain second order spatial derivatives. Among
the operators of dimension higher than four, the simplest possibility that
gives the correction to the contact scattering, the $\lambda ^2$ term in (%
\ref{ALsubdom}), can be borrowed from our earlier calculation of the
nonrelativistic limit of $\phi ^4$ theory \cite{go}. In fact, the
addition of the interaction Lagrangian $\frac{v_1}{4 m^2}\left[ \psi ^{*}(%
{\bf \nabla }^2\psi ^{*})\psi ^2-\left( {\bf \nabla }\psi ^{*}\right) ^2\psi
^2\right] $ produces, for the one-loop CM scattering, a term proportional to 
${\bf p}^2\left[ \ln (\Lambda ^2/{\bf p}^2)+i\pi \right] $ so that the first
parcel of (\ref{ALsubdom}) can be obtained by properly adjusting the value
of the parameter $v_1$. Similarly, one can easily sees that a seagull
derivative interaction of the kind $\frac{c_1}{8m^3}e^2{\bf A}^2\psi ^{*}%
{\bf \nabla }^2\psi $ generates the parcel of (\ref{ALsubdom}) proportional
to $e^4$ by adequately choosing the value of $c_1$. These matchings furnish $%
v_1=-5\lambda / 64 m^2$ and $c_1=-1$ .

These additional interactions should, of course, be modified by taking
covariant derivatives instead of ordinary ones, in order to guarantee gauge
invariance. One is thus lead to suggest the following effective interaction

\begin{equation}
{\cal L}_{int}^{NR}=-\frac{5\lambda}{256m^4}\,\left[ \psi ^{*}{\bf D}^2\psi
^{*}+\left( {\bf D}\psi ^{*}\right) ^2\right] \,\psi ^2\,-\,\psi ^{*}%
\frac{{\bf D}^4}{8m^3}\psi \;\;\;.  \label{Leff}
\end{equation}
Notice that the new terms that appear by considering covariant derivatives
do not contribute up to the order we have worked, since they involve powers
higher than two of either $|{\bf p}|$ or $e\,$. The last term in the above
expression can be seen as originated from the expansion of the energy
dispersion relation like that which happens in the Foldy-Wouthuysen
procedure. There are other dimension six operators which could also have
been considered. For example, by extending the covariant derivative in a
non-minimal way through the introduction of a magnetic term, that is by
defining ${\cal D}_j=D_j+e\,\epsilon _{j\sigma \rho }\partial ^\sigma A^\rho
\,$, the $-\frac 1{2m}({\cal D}\psi )^{*}\cdot ({\cal D}\psi )\,$ term
in the Lagrangian introduces (besides the usual $|\,{\bf D}\psi\,|^{2}\,$)
parcels involving ${\bf E\times }\left( \psi ^{*}{\bf %
\nabla }\psi \right) \,\,$, $\,{\bf E\times A\,}\psi ^{*}\psi \,\,$ and
\thinspace ${\bf E}^2\,\psi ^{*}\psi \,$, which might contribute to the
scattering amplitude at order ${\bf p}^2/m^2\,$. However, to take into
account appropriately all the possibilities and to derive a more complete
effective Lagrangian, one has to consider other sectors of the theory. This
analysis is left for future work.

\section{Appendix}

We summarize here the calculation of the various contributions to the CS
1-loop particle-particle scattering in the CM\ frame, using the
approximation described in Sec. III. The momentum assignment is that of Fig.
8.

\noindent $(i)$ {\bf Direct box graph}.

Following the Feynman rules, one has

\begin{eqnarray}
A^{(b)dir} &=&-ie^4\int \frac{d^3k}{(2\pi )^3}\left\{ (p_1+k)^\mu \,D_{\mu
\sigma }(k-p_1)\,(2p_2+p_1-k)^\sigma \,\Delta (k)\right.  \nonumber \\
&&\ \ \left. \Delta (p_1+p_2-k)\,(-k+p_1+p_2+p_2^{\prime })^\rho \,D_{\rho
\nu }(k-p_1^{\prime })\,(k+p_1^{\prime })^\nu \right\} \,+\,\left[
p_1^{\prime }\leftrightarrow p_2^{\prime }\right]  \nonumber \\
\ &=&-\frac{4e^4}{\pi ^2\Theta ^2}\int d^2{\bf k\,}\left( \frac{w_p^2}{w_k}%
\right) \,\frac 1{{\bf p}^2-{\bf k}^2+i\varepsilon }\,\left[ \frac{({\bf %
k\times p})\,({\bf k\times p}^{\prime })}{({\bf k-p})^2({\bf k-p}^{\prime
})^2}\right] \,+\,\left[ {\bf p}^{\prime }\leftrightarrow -{\bf p}^{\prime
}\right] \;,  \label{abd}
\end{eqnarray}
where the $k^0$ integration of the CM amplitude was done as a contour
integral. The angular integration, which involves only the third factor of
the integrand in the last line above, can be cast in the form $\frac 12%
\left[ \cos \theta \,{\rm I}_0-{\rm I}_2\right] $ where

\begin{equation}
{\rm I}_n=\int_0^{2\pi }d\varphi \,\frac{\cos \,(n\varphi )}{\left[ 2\cos
(\varphi -\theta /2)-\beta \right] \left[ 2\cos (\varphi +\theta /2)-\beta
\right] }  \label{In}
\end{equation}
and $\beta =({\bf k}^2+{\bf p}^2)/(|{\bf k}||{\bf p}|)$ . Making $z=\exp
(i\varphi )$ and using the residue theorem, one finds

\begin{equation}
{\rm I}_n=\frac \pi {{\rm B}\sqrt{\beta ^2-4}}\frac{(\vartheta
^{n-1}+\vartheta ^{-n+1})\sin \left[ (n+1)\theta /2\right] -(\vartheta
^{n+1}+\vartheta ^{-n-1})\sin \left[ (n-1)\theta /2\right] }{\sin (\theta /2)%
}  \label{Inf}
\end{equation}
where ${\rm B}=\beta ^2-2(1+\cos \theta )$ and $\vartheta =\frac 12\left[
\beta -\sqrt{\beta ^2-4}\right] $ . Using this formula, one can write

\begin{eqnarray}
A^{(b)dir} &=&-\frac{e^4}{\pi \Theta ^2}\int d({\bf k}^2)\ \left( \frac{w_p^2%
}{w_k}\right) \frac 1{{\bf p}^2-{\bf k}^2+i\varepsilon }  \nonumber \\
&&\hspace{2.5cm}\left[ \frac{|\,{\bf k}^2-{\bf p}^2\,|\,(\,{\bf k}^2+{\bf p}%
^2\,)}{({\bf k}^2)^2+({\bf p}^2)^2-2\,{\bf k}^2\,{\bf p}^2\,\cos \theta }%
-1\right] \,+\,\left( \theta \leftrightarrow \pi +\theta \right) \;.
\label{abd2}
\end{eqnarray}
The remaining ${\bf k}^2$ integration is then divided into two pieces, from $%
0$ to $\Lambda ^2$ (L region) and from $\Lambda ^2$ to $\Lambda
_0^2\rightarrow \infty $ (H sector). In the L part, using

\begin{equation}
\frac{w_p^2}{w_k}=m\left( 1+\frac{{\bf p}^2}{m^2}\right) \left[ 1-\frac{{\bf %
k}^2}{2m^2}+\frac{3({\bf k}^2)^2}{8m^4}+...\right]  \label{wpwk}
\end{equation}
and keeping terms up to order $\eta ^2$, one obtains

\begin{eqnarray}
A_L^{(b)dir} &\simeq &-\frac{e^4}{\pi \Theta ^2}m\left\{ \left( 1+\frac{{\bf %
p}^2}{2m^2}\right) \left[ \ln (2\left[ 1-\cos \theta \right] )+i\pi \right]
-(1-\cos \theta -2\cos ^2\theta )\frac{{\bf p}^4}{\Lambda ^4}-\frac{{\bf p}^2%
}{m^2}\right.  \nonumber \\
&&\ \ \left. -\cos \theta \,\left[ \ln (2\left[ 1-\cos \theta \right] )+\ln
\left( \frac{{\bf p}^2}{\Lambda ^2}\right) \right] \,\frac{{\bf p}^2}{m^2}%
+2\cos \theta \frac{{\bf p}^2}{\Lambda ^2}\right\} +\left( \theta
\leftrightarrow \pi +\theta \right) \ .  \label{abdL}
\end{eqnarray}
Since $\cos (\pi +\theta )=-\cos \theta $ , one recovers equation (\ref
{AbdirL}). In the H region, the integrand is replaced by its Taylor
expansion around ${\bf p}^2=0$ which is given by 
\[
\left[ -\frac{2m^2\cos \theta }{({\bf k}^2)^2\sqrt{{\bf k}^2+m^2}}\right] 
{\bf p}^2+\left[ \frac{2m^2(1-2\cos ^2\theta )}{({\bf k}^2)^3\sqrt{{\bf k}%
^2+m^2}}-\frac{2\cos \theta \sqrt{{\bf k}^2+m^2}}{({\bf k}^2)^3}\right] {\bf %
p}^4+{\cal O}({\bf p}^6). 
\]
Performing the ${\bf k}^2$ integrations one obtains, up to order $\eta ^2$,

\begin{eqnarray}
A_H^{(b)dir} &\simeq &-\frac{e^4}{\pi \Theta ^2}m\left\{ -\cos \theta \ln
\left( \frac{\Lambda ^2}{4m^2}\right) \frac{{\bf p}^2}{m^2}-2\cos \theta 
\frac{{\bf p}^2}{\Lambda ^2}\right.  \nonumber \\
&&\left. \hspace{1.8cm}+(1-\cos \theta -2\cos ^2\theta )\frac{{\bf p}^4}{%
\Lambda ^4}-\cos \theta \frac{{\bf p}^2}{m^2}\right\} +\left( \theta
\leftrightarrow \pi +\theta \right) \ ,  \label{abdh}
\end{eqnarray}
which is the same as (\ref{AbdirH}).

\noindent $(ii)$ {\bf Twisted box diagram}.

After performing the $k^0$ integration, one gets

\begin{eqnarray}
A^{(b)twist} &=&-\frac{e^4}{2\pi ^2\Theta ^2}\int d^2{\bf k\,}\left( \frac{%
w_kw_{k-s}-w_p^2}{w_kw_{k-s}(w_k+w_{k-s})}\right)  \nonumber \\
&&\ \hspace{2.5cm}\left[ \frac{({\bf k\times s})^2-({\bf p\times p}^{\prime
})^2}{({\bf k-p})^2({\bf k-p}^{\prime })^2}\right] \,+\,\left[ {\bf p}%
^{\prime }\leftrightarrow -{\bf p}^{\prime }\right]  \label{abt}
\end{eqnarray}
where ${\bf s=p}+{\bf p}^{\prime }$ and we recall that $w_{k-s}=\sqrt{({\bf k%
}-{\bf s})^2+m^2}$ . The angular integration above is very complicated and,
therefore, we separate the L and H sectors before make it. The $1/m$
expansion of the first factor in the integrand of (\ref{abt}) is

\[
\frac 12\left( {\bf k}^2+\cos \theta \,{\bf p}^2-\sqrt{2(1+\cos \theta )}|%
{\bf k}|\,|{\bf p}|\,\cos \varphi \right) \frac 1{m^3}+{\cal O}(1/m^5)\;. 
\]
One immediately suspect that there will be no contribution from the L sector
up to order $\eta ^2$. Pushing further, one can do the angular integration,
which is expressible in terms of ${\rm I}_n\,$ $(n=0,...,3)$, and the ${\bf k%
}^2$ integration, to get up to order $\eta ^3$,

\begin{equation}
A_L^{(b)twist}\approx -\frac{e^4}{4\pi \Theta ^2}m(1+\cos \theta )\frac{{\bf %
p}^2\Lambda ^2}{m^4}+\left( \theta \leftrightarrow \pi +\theta \right) \ .
\label{abtl}
\end{equation}
In the H region, the whole integrand is expanded around $|{\bf p}|=0$
resulting in

\[
\left( \frac{(1+\cos \theta )\sin ^2\varphi }{({\bf k}^2+m^2)^{3/2}}\right)
\,{\bf p}^2+{\cal O}(|{\bf p|}^3)\;, 
\]
and, thus, the integrations over $\varphi $ and ${\bf k}^2$ lead to

\begin{equation}
A_H^{(b)twist}\approx -\frac{e^4}{2\pi \Theta ^2}m\left\{ (1+\cos \theta )%
\frac{{\bf p}^2}{m^2}-\frac 12(1+\cos \theta )\frac{{\bf p}^2\Lambda ^2}{m^4}%
\right\} +\left( \theta \leftrightarrow \pi +\theta \right) \ ,  \label{abth}
\end{equation}
which reproduces (\ref{AbtwistLH}).

\noindent $(iii)$ {\bf Seagull scattering}.

The gauge bubble contribution to the amplitude in the CM frame is given by

\[
-\frac{4e^4}{\Theta ^2}\int \frac{d^3k}{(2\pi )^3}\,\frac{({\bf k}-{\bf p}%
)\cdot ({\bf k}-{\bf p}^{\prime })}{({\bf k-p})^2({\bf k-p}^{\prime })^2}\;. 
\]
The linear divergence of this expression manifests itself at once in the $%
k^0 $ integration. However, the sum of the equal triangle diagrams has a
parcel with a divergent $k^0$ integration which exactly cancel the preceding
one so that one can write

\begin{eqnarray}
A^{(c)} &=&-\frac{e^4}{2\pi ^2\Theta ^2}\left\{ \int d^2{\bf k}\left( \frac{%
w_p^2+w_k^2}{w_k}\right) \,\frac{({\bf k}-{\bf p})\cdot ({\bf k}-{\bf p}%
^{\prime })}{({\bf k-p})^2({\bf k-p}^{\prime })^2}\right.  \nonumber \\
&&\left. \hspace{1.8cm}-\int d^2{\bf k\,}\left( \frac 4{w_k}\right) \left[ 
\frac{({\bf k\times p})\,({\bf k\times p}^{\prime })}{({\bf k-p})^2({\bf k-p}%
^{\prime })^2}\right] \right\} \,+\,\left( \theta \leftrightarrow \pi
+\theta \right) \ .  \label{ac}
\end{eqnarray}
The angular integration of the second term above is the same as that of the
direct box diagram while that of the first parcel can be written as a linear
combination of ${\rm I}_0$ and ${\rm I}_1$, and doing so, one gets

\begin{eqnarray}
A^{(c)} &=&-\frac{e^4}{2\pi \Theta ^2}\left\{ \int d({\bf k}^2)\,\left( 
\frac{w_p^2+w_k^2}{w_k}\right) \,\frac{{\rm sgn}({\bf k}^2{\bf -p}^2)\,(\,%
{\bf k}^2-\cos \theta \,{\bf p}^2\,)}{({\bf k}^2)^2+({\bf p}^2)^2-2\,{\bf k}%
^2\,{\bf p}^2\,\cos \theta }\right.  \nonumber \\
&&\ \ \left. \hspace{1.0cm}-\int d({\bf k}^2){\bf \,}\frac 1{w_k}\,\left[ 
\frac{|\,{\bf k}^2-{\bf p}^2\,|\,(\,{\bf k}^2+{\bf p}^2\,)}{({\bf k}^2)^2+(%
{\bf p}^2)^2-2\,{\bf k}^2\,{\bf p}^2\,\cos \theta }-1\right] \right\}
\,+\,\left( \theta \leftrightarrow \pi +\theta \right) \ ,  \label{acf}
\end{eqnarray}
where ${\rm sgn}$ denotes the signal function. Using (\ref{expanwk}) and (%
\ref{wpwk}) and performing the integrations, the L sector contribution is
obtained, up to order $\eta ^2$, as

\begin{eqnarray}
A_L^{(c)} &\simeq &-\frac{e^4}{2\pi \Theta ^2}m\left\{ \left( 2+\left[
1-2\cos \theta \right] \frac{{\bf p}^2}{m^2}\right) \left[ \ln \left( \frac{%
\Lambda ^2}{{\bf p}^2}\right) -\ln (2\left[ 1-\cos \theta \right] |)\right]
\right.  \nonumber \\
&&\ \ \ \left. \hspace{1.5cm}+2\frac{{\bf p}^2}{m^2}-2\cos \theta \frac{{\bf %
p}^2}{\Lambda ^2}+(1-2\cos ^2\theta )\frac{{\bf p}^4}{\Lambda ^4}+\frac{%
\Lambda ^4}{8m^4}\right\} \,+\,\left( \theta \leftrightarrow \pi +\theta
\right) \ ,  \label{acl}
\end{eqnarray}
which coincides with (\ref{AcL}). The H part, cutoff regulated, is obtained
by expanding the integrand up to ${\bf p}^4$ and integrating from $\Lambda
^2 $ to $\Lambda _0^2$. It is found

\begin{eqnarray}
A_H^{(c)} &\simeq &-\frac{e^4}{2\pi \Theta ^2}m\left\{ -\left( 2+\left[
1-2\cos \theta \right] \frac{{\bf p}^2}{m^2}\right) \ln \left( \frac{\Lambda
^2}{4m^2}\right) -2+2\cos \theta \frac{{\bf p}^2}{\Lambda ^2}\right. 
\nonumber \\
&&\ \ \ \ \left. \hspace{1.4cm}+\cos \theta \,\frac{{\bf p}^2}{m^2}-(1-2\cos
^2\theta )\frac{{\bf p}^4}{\Lambda ^4}-\frac{\Lambda ^4}{8m^4}\right\} \,-%
\frac{e^4}{\pi \Theta ^2}\,\Lambda _0+\,\left( \theta \leftrightarrow \pi
+\theta \right) \ ,  \label{ach}
\end{eqnarray}
the same as (\ref{AcH}). 
\begin{flushleft}
{\bf Acknowledgements}
\end{flushleft}
This work was partially supported by Conselho Nacional de Desenvolvimento
Cient\'\i fico e Tecnol\'ogico (CNPq) e Funda\c c\~ao de Amparo \`a Pesquisa
do Estado de S\~ao Paulo (FAPESP).

\newpage

\begin{center}
Figure captions
\end{center}

\vspace{1.0cm}

Fig.1 - Feynman rules for the interaction vertices.

\vspace{1.0cm}

Fig. 2 - Particle self-energy insertion in 1-loop order.

\vspace{1.0cm}

Fig. 3 - Vacuum polarization correction in one loop order.

\vspace{1.0cm}

Fig. 4 - One loop correction to the trilinear vertex.

\vspace{1.0cm}

Fig. 5 - One loop correction to the seagull vertex.

\vspace{1.0cm}

Fig. 6 - Tree level scattering.

\vspace{1.0cm}

Fig. 7 - Basic mixing interactions diagrams .

\vspace{1.0cm}

Fig. 8 - One loop order particle-particle scattering. In the momenta
assignment shown, which is used in the calculations, $s=p_1+p_2$ , $%
q=p_1-p_1^{\prime }$ and $u=p_1^{\prime }-p_2$ .


\begin{references}
\bibitem{f1}   The origin of each term can be
traced back and it is found that the finite relativistic scattering has
contributions coming from both L and H sectors. This shows that the
claim of ref. \cite{boz} that the divergence of the
2-particle NR amplitude is due to the neglection of parts of the diagrams
involving pair production is misleading. As demonstrated in ref. \cite{go}, 
even if one completely excludes the part that represents propagation
backwards in time from the relativistic propagator, the fish amplitude
remains finite.
\end{references}
\end{document}